\documentclass[aps,prc,twocolumn,superscriptaddress,nofootinbib,showpacs]{revtex4}

\usepackage{amsmath}
\usepackage{graphicx}

\newcommand{\re}{\ref}
\newcommand{\be}{\begin{equation}}
\newcommand{\ee}{\end{equation}}

\newcommand{\la}{\label}
\newcommand{\ber}{\begin{eqnarray}}
\newcommand{\eer}{\end{eqnarray}}

\begin{document}

\sloppy

\title{Properties of the first excited state of $^9$Be derived from $(\gamma,n)$ and ($e,e'$) reactions}


\newcommand{\Kurchatov}{National Research Center  "Kurchatov Institute",  123182 Moscow,  Russia}
\newcommand{\TUDarmstadt}{Institut f\"ur Kernphysik, Technische Universit\"{a}t Darmstadt, D-64289 Darmstadt, Germany}
\newcommand{\ECT}{European Centre for Theoretical Studies in Nuclear Physics and Related Areas (ECT*), Villa Tambosi, I-38123 Villazzano (Trento), Italy}

\author{V.~D.~Efros}\email{v.efros@mererand.com}\affiliation{\Kurchatov} \affiliation{\ECT}
\author{P.~von~Neumann-Cosel}\affiliation{\TUDarmstadt}
\author{A.~Richter}\affiliation{\ECT}\affiliation{\TUDarmstadt}

\date{\today}

\begin{abstract}
Properties of the first excited state of the nucleus $^9$Be are discussed based on recent $(e,e')$ and $(\gamma,n$) experiments. 
The parameters of an $R$--matrix analyses of different data sets are consistent with a resonance rather than a virtual state 
predicted by some model calculations. 
The energy and the width of the resonance are deduced. 
Their values are rather similar for all data sets, and the energy proves to be negative. 
It is argued that the disagreement between the extracted $B(E1)$ values may stem from different ways of integration of the resonance.
If corrected, fair agreement between the $(e,e')$ and one of the  $(\gamma,n)$ data sets is found.
A recent $(\gamma,n)$ experiment at the HI$\gamma$S facility exhibits larger cross sections close to the neutron threshold which remain 
to be explained.
\end{abstract}

\pacs{21.10.Tg, 25.20.-x, 25.30.Dh, 27.20.+n}

\maketitle

\section{Introduction}

The first excited state in $^9$Be with quantum numbers $J^\pi = 1/2^+$ lies in the vicinity of the neutron emission threshold. 
It is of special interest to understand the  three--cluster structure of this state, and in particular the $^9$Be photodisintegration 
via this state. 
The inverse fusion process is of importance in explosive nucleosynthesis scenarios providing seed material for a production of $^{12}$C 
via the ($\alpha \alpha n)^9{\rm Be}(\alpha,n)^{12}{\rm C}$ reaction chain  alternative to the triple--alpha reaction \cite{sasa}.

Therefore in recent years several studies of this state  with electromagnetic probes were performed  using both $(\gamma,n)$ and $(e,e')$ reactions.
$R$--matrix parameters and the $E1$ strength of the transition to this state from the ground state  were determined \cite{darms,sum,arnold}. 
(Note that the analysis in Ref.~\cite{sum} includes the data of Ref.~\cite{jap}.) 
The consistency with previous work is discussed in Refs.~\cite{darms,arnold}.
Since the state lies so close to the threshold  the extraction of its properties is non--trivial.
The $B(E1)$ values reported in Refs. \cite{darms,sum,arnold} differ significantly from each other.
In the present Brief Report a possible explanation is offered, and after the 
reinterpretation the $(e,e')$ and one of the $(\gamma,n)$ data sets are in good agreement.
Using the experimental data we also determine the energy of the $1/2^+$ state.

\section{Energy of the first excited state of ${\bf^9}{\rm\bf Be}$}

In this section the energy of the $1/2^+$ state is extracted from the recent $(\gamma,n)$ and $(e,e')$ data.
One point discussed in the literature is whether the first excited state (1/2$^+$) of the $^9$Be nucleus is a resonance or a virtual state. 
An answer to the question would provide both a constraint on nuclear forces to be used in microscopic calculations and an accuracy test of calculations.

A virtual state or a resonant state reveals itself as a pole in the  $S$ matrix $S_{\alpha\beta}(E)$.
In the resonant state case the $S$ matrix pole is of the form \mbox{$[E-(E_{res}-i\Gamma/2)]^{-1}$} with $\Gamma>0$.
The virtual state case is to be considered within the approximation that $^8{\rm Be}$ is a stable particle and $^8$Be-$n$ is 
thus the lowest $^9{\rm Be}$ threshold. 
In this case the pole dependence turns to $(E-E_0)^{-1}$ with $E_0$ being real and negative. 
In microscopic calculations, methods dealing with energies away from the scattering line are applied to obtain $E_{res}$ and 
$\Gamma$, or  $E_0$ \cite{arai,koike,arai1,Furutani,gar,dima}. 
Both a virtual \cite{koike,arai1} and a resonant 1/2$^+$ state \cite{Furutani,gar,dima} emerged in the calculations. 

In the experiments of Refs. \cite{darms,jap,sum,arnold} the photoabsorption  cross section for the transition of the $^9$Be 
nucleus from its ground state to the first excited state was studied. 
After separating out the contributions of tails of other resonances the cross section data were fitted with an expression of the form
\be 
\sigma(E_\gamma)=\frac{16\pi^2}{9}\frac{e^2}{\hbar c}C
\frac{E_\gamma\sqrt{\epsilon(E_\gamma-S_n)}}{(E_\gamma-E_R)^2+\epsilon(E_\gamma-S_n)}.
\la{crs}
\ee
Here \mbox{$S_n=1.6654$ MeV} is the $^8$Be-$n$ threshold, $E_\gamma$ is the photon energy, 
and $C$ is a constant discussed below.
The notation \mbox{$\epsilon =(\Gamma_R/2)^2/(E_R-S_n)$} is used, and $E_R$ and $\Gamma_R$ are fitting parameters. 
Eq.~(\re{crs}) implies the decay of the first excited state into $^8$Be+$n$ and employs the one--level \mbox{$R$--matrix} approximation.
With these assumptions such a representation of the cross section  is valid irrespective of the structure of the 1/2$^+$
state in the inner region.
In fact Eq.~(\re{crs}) includes the correct form of a pole contribution to the energy dependence of  the cross section in the threshold region    
independently  of the \mbox{$R$--matrix} approach. 

In the further discussion it is convenient to use the quantities \mbox{${\bar E_R}=E_R-S_n$} and \mbox{$E_x=E_\gamma-S_n$}, i.e., 
the \mbox{$R$--matrix} energy defined with respect to the threshold and the excitation energy instead of the photon energy. 
Correspondingly,  the energy dependence of Eq.~(\re{crs}) takes the form  
\be 
\sigma \propto \frac{(E_x+S_n)\sqrt{\epsilon E_x}}{[(E_x-{\bar E}_R)^2+\epsilon E_x]}.
\la{ende}
\ee
Expression (\re{ende}) includes a pole corresponding to the $1/2^+$ state in the spectrum of $^9$Be, which may be a resonant one or a virtual one. 
While there also exist non--pole contributions to the cross section it is assumed here that the pole is located sufficiently 
close to the scattering line so that its contribution is dominating. 
For the determination of the pole position it is then required to represent the denominator 
of the expression (\re{ende}) in terms of its roots, i.e. as \mbox{$(E_x-E_{1})(E_x-E_{2})$}. 
If the state is a resonant one then \mbox{$E_{1,2}=E_{res}\pm i\Gamma/2$}. 
For a virtual state $E_{1,2}$ are real and negative.  

Let us find out the nature and the position of the pole based on the $R$--matrix parameters of the experiments 
listed in the first and second column of the Table \ref{table1}.
Considering the roots $E_{1,2}$ one sees that a state is resonant when \mbox{$\Gamma_R/{\bar E}_R<4$}  and virtual one otherwise.
Thus, the results of Tab.~\ref{table1} are consistent with a resonant state.
\begin{table}
\caption{$R$-matrix parameters (${\bar E}_R$,$\Gamma_R$) 
extracted in recent experiments using real and virtual photons and the corresponding positions ($E_{res}$) and the widths ($\Gamma$)
of the $^9$Be $1/2^+$ resonant state. 
\label{table1}} 
\begin{center}
\begin{tabular}{ccccccc}
\hline \hline
Ref. & ${\bar E}_R$ & $\Gamma_R$ & $E_{res}$  & $\Gamma$\\
   &  (keV)  & (keV)& (keV) & (keV)\\
\hline
\cite{darms} & 83(6) & 274(8) &  -31(6) & 153(4) \\
\cite{sum} & 70(3) & 225(12) & -21 & 132 \\
\cite{arnold} & 66(2) & 213(6) & -21 & 124 \\
\hline
\end{tabular}
\end{center}
\end{table} 

The position $E_{res}$ and the width $\Gamma$ of the resonance are listed in the third and fourth columns of the Table for each set of the 
experimental \mbox{$R$--matrix} parameters ${\bar E}_R$ and $\Gamma_R$. They are related by
\[ E_{res}={\bar E}_R-\epsilon/2,\qquad
\Gamma=2(\epsilon{\bar E}_R-\epsilon^2/4)^{1/2}.\]
A peculiar point is that the $E_{res}$ values obtained are negative while one typically has $E_{res}\gg\Gamma>0$. 
One sees that
a value $E_{res} > 0$  would have been obtained if the condition \mbox{$\Gamma_R/{\bar E}_R\le2\sqrt{2}$} was fulfilled.
The uncertainties of the $E_{res}$ and $\Gamma$ parameters deduced from the Darmstadt data \cite{darms} were obtained by
fitting the spectrum (top part of Fig. 5 in Ref. \cite{darms}) using Eq.~(\re{ende}) with the denominator being written as \mbox{$(E_x-E_{res})^2+(\Gamma/2)^2$}.
A $\chi^2=0.64$ value per datum was obtained in this fit. 

The negative resonance energy can be interpreted in the following way.
It is a resonance of the three--body $\alpha$-$\alpha$-$n$ structure
in the inner region. 
Since the decay of the $^8$Be ground state into two $\alpha$ particles is unbound by 91 KeV,  the resonance energy is positive with respect to the $\alpha$-$\alpha$-$n$ threshold. 
The three--body decay of the resonance is strongly suppressed due to the Coulomb barrier between the $\alpha$ particles so 
that it decays via the $^8$Be-$n$ channel. 
Its width is narrow because the overlap with the $^8$Be-$n$ cluster structure is small.
The energy difference between the $^8$Be-$n$ threshold ($S_n$) and $E_{res}$ is still smaller than the half width of the resonance
leading to the pronounced peak in the $(\gamma,n)$ cross section above $S_n$.

The true cross section includes non--resonant contributions not present in Eqs.~(\re{crs}) and (\re{ende}) that may influence
the ($E_{res}$, $\Gamma$) parameters.  
For an estimate of this effect let us model the energy dependence of the transition probability by the expression 
$\int \psi_k^2 d\tau_{(\rho)}$ where  $(\hbar k^2)/(2\mu)=E_x$ and $\psi_k$ is the $s$--wave component of the \mbox{$^8$Be-$n$} 
final state wave function normalized to $\delta({\bf k}-{\bf k}')$.
The integration is extended over the interaction region of the range of $\rho$. 
A value $\rho=4$~fm corresponding to the sum of the root--mean--square radius of the $^8$Be matter distribution and 
the nucleon--nucleon interaction range is assumed. 

The resulting energy dependence \cite{lud}, applicable in the general many--body one--channel case, is universal. (It is
obtained setting \mbox{$\chi_{k}(r)=r\psi_{k}(r)$}, writing the quantity \mbox{$\int_0^{\rho}dr\chi_{k_1}(r)\chi_{k}(r)$} as
\mbox{$(E_1-E)^{-1}[-\hbar^2/(2\mu)][\chi_{k_1}(\rho)\chi'_{k}(\rho)-\chi_{k}(\rho)\chi'_{k_1}(\rho)]$}, 
and taking the limit $E_1\rightarrow E$.) 
The corresponding  cross section may be written as
\begin{widetext}
\be 
(E_x+S_n)k \int \psi_k^2 d\tau_{(\rho)} \propto \frac{E_x+S_n}{k} 
\left\{\left[\frac{1}{2i}\frac {d\ln S(k)}{dk}-\frac{\cos2k\rho}{2k}
{\rm Im}S(k)\right]+\rho\left[1-\frac{\sin2k\rho}{2k\rho}{\rm Re}S(k)\right]\right\},
\la{ed}
\ee
\end{widetext}
where $S(k)$ is the $S$-matrix element for \mbox{$^8$Be-$n$} scattering.
It is of the form \mbox{$[g(k^2) + ik]/[g(k^2) - ik]$}, where $g(k^2)$ is real. 
At the low energies under consideration, $g(k^2)$ may be estimated in the effective range approximation \mbox{$g(k^2)\simeq-1/a+(r_0/2)k^2$}.
If the $S$ matrix includes a resonant pole, the corresponding effective range approximation expression may be rewritten as 
\be 
S(k)=\frac{(k-k_0^*)(k+k_0)}{(k-k_0)(k+k_0^*)}
\la{s}
\ee 
where \mbox{$(\hbar k_0^2)/(2\mu)=E_{res}-i\Gamma/2$}, \mbox{$k_0=k_1-ik_2$}, and \mbox{$k_{1,2}>0$}.  
Since in our case \mbox{$E_{res}<0$}, one has \mbox{$k_2>k_1$}. 
Expression (\re{ed}) includes both a pole and a non--pole energy dependence.

Let us perform a least--square fit of Eq.~(\re{ed}) assuming the same energy dependence as in Eq.~(\re{ende})
\be 
(E_x+S_n)\sqrt{E_x}/[(E_x-E_{res}^{mod})^2+(\Gamma_{mod}/2)^2].
\la{efit}
\ee 
Here $E_{res}^{mod}$ and $\Gamma_{mod}$ are fitting parameters.
Taking, e.g., the values \mbox{$(E_{res},\Gamma_{res}) = (-31,153)$}~keV in the Table
deduced from the data of Ref.~\cite{darms}, 
parameters \mbox{$(E_{res}^{mod},\Gamma_{mod})=(-20, 146)$~keV} are obtained.
For \mbox{$(E_{res},\Gamma_{res})=(-21,124)$}~keV obtained from the data of Ref.~\cite{arnold}, 
the corresponding parameters \mbox{$(E_{res}^{mod},\Gamma_{mod})$} are \mbox{(-13, 119)}~keV. 
The fit was performed in the energy interval $0\le E_x\le 0.54$ MeV which just covers the $1/2^+$ peak. 
The average relative deviation between Eqs.~(\re{ed}) and  (\re{efit}) in this energy region is about 2\%.
Variations of $\rho$ within 1~fm lead  to negligible effects on $E_{res}^{mod}$ and $\Gamma_{mod}$. 
The differences between the sets \mbox{$(E_{res},\Gamma_{res})$} and \mbox{$(E_{res}^{mod},\Gamma_{mod})$} 
are reasonably small in all the cases indicating that the quantities \mbox{$E_{res}$} and $\Gamma_{res}$  in  the Table  
are only weakly affected by non--resonant contributions to the cross section.

In Ref.~\cite{bang} it was deduced that the $^9$Be 1/2$^+$ state is a virtual state. 
However,  the analysis \cite{bang} relied on the rather limited photodisintegration cross section data prior to the measurements 
of Refs.~\cite{sum,arnold}, and the influence of experimental uncertainties was not considered. 
In the nine--nucleon dynamics calculation of Ref.~\cite{arai} no 1/2$^+$ state was detected at all. 
A possible reason is that the NN forces used lead to a virtual 1/2$^+$ state while resonant states, i.e.\ complex--energy eigenstates, 
were being searched for in the calculation. 
The 1/2$^+$ state was detected as a virtual state in the $\alpha$-$\alpha$-$n$--model calculation \cite{koike} and in the nine--nucleon 
dynamics calculation \cite{arai1}. 
In the nine--nucleon resonating group model it was assumed that the 1/2$^+$ state is a resonant state and its width $\Gamma$ was 
found to be 91 keV \cite{Furutani}. 

In the $\alpha$-$\alpha$-$n$--model calculation of Refs.~\cite{gar,dima} the parameters of the resonance were determined to be
\mbox{$E_{res}=(+)18.6$ keV} and \mbox{$\Gamma\simeq100$ keV}.
The small width was explained by the fact that the state has a \mbox{$^5$He+$\alpha$} structure at small distances different 
from its \mbox{$^8$Be+$n$} structure at large distances. 
It is commonly adopted that the amount of the decay of the 1/2$^+$ state via the \mbox{$^8$Be-$n$} channel is about 100\% 
(see, e.g., \cite{arai1,dima}). 
In addition, as shown above the non--resonant contributions to the cross section are small.
Taking this into account the applicability of the one--channel $R$ matrix representation of the cross section is not affected 
by the mentioned change of the cluster structure with relative distance. Therefore there is no reason to attribute 
differences between experimental and theoretical widths to use of the  \mbox{$R$--matrix} representation.
Moreover, as mentioned above Eq.~(\ref{crs}) is independent of the $R$--matrix parametrization. 

\section{$B(E1)$ strength of the transition to the 1/2$^+$ state} 

Here the data on the $B(E1)$ strength for the $(\gamma,n)$ transition to  the 1/2$^+$ state are analyzed along the lines of 
Refs.~\cite{clerc,barker1}. 
Both the $B(E1)$ strengths obtained  in $(\gamma,n)$ experiments \cite{sum,jap,arnold} and deduced from $(e,e')$ experiments 
\cite{darms,bates,darms1} are considered.

The photo--absorption cross section for the excitation of the 1/2$^+$ state
may be written as
\be
\sigma(E_\gamma)=\frac{16\pi^3}{9}\frac{e^2}{\hbar c} E_\gamma b(E1,E_\gamma)\la{sg}
\ee
where $b(E1,E_\gamma)$ is the  reduced transition strength distribution. Here 
the contributions to the cross section 
of states different from the \mbox{1/2$^+$ state} 
are assumed to be subtracted, 
and non--resonant contributions are disregarded. 
The transition is caused by the $E1$ multipole of the current, and in the long--wave Siegert approximation the quantity  $b(E1,E_\gamma)$ 
may be represented in the form of the unretarded $C1$ strength distribution. 
This makes it possible to relate this quantity to the $(e,e')$ longitudinal response function in the low momentum--transfer limit, i.e. 
to deduce the cross section (\re{sg}) from ($e,e'$) experiments.
In case of the usual one--body charge operator the corresponding $C1$ expression is \cite{ajz}
\begin{widetext}
\be b(E1,E_\gamma)=\frac{1}{2J_i+1}\sum_{M_i=-J_i}^{J_i}\sum_{\mu=\pm1,0}
\int df
\left|\left[\sum\nolimits_{j=1}^Zr_j'Y_{1\mu}({\bf r}_j'/r_j')\right]_{fi}\right|^2
\delta(E_f-E_i-E_\gamma).
\la{be1}
\ee 
\end{widetext}
Here $i$ and $f$ refer to an initial state and to final states, and 
${\bf r}_j'$ is a proton position with respect to the center of mass of a system. 

The integral reduced  strength  $B(E1)\!\!\uparrow$ of a level is defined as an integral  "over the peak" of the reduced strength distribution,
\be
B(E1)\!\!\uparrow=\int_{S_n}^{E_{max}}b(E1,E_\gamma)dE_\gamma.\la{BE1}
\ee
In the analysis of the $(\gamma,n)$ experiments \cite{sum,jap,arnold} the fitting  constant $C$ from Eq. (\re{crs}) was referred to as the  
reduced strength for the transition to the \mbox{1/2$^+$ state,}
\be 
C=B(E1)\!\!\uparrow.
\la{def}
\ee
Equation~(\ref{def}) can be derived from Eq.~(\ref{BE1}) by integration up to $E_{max} = \infty$.
For any  ${\bar E}_R$ and $\epsilon>0$ values one has 
\be
\int_0^{\infty}\frac{\sqrt{\epsilon E_x}}{(E_x-{\bar E}_R)^2+\epsilon E_x}dE_x=\pi,
\la{pi}
\ee
which leads directly to Eq.~(\re{def}). 

The values thus obtained from the $(\gamma,n)$ data for the quantity (\re{def}) were 
\mbox{$0.052(1)$~fm$^2$~\cite{sum}}, \mbox{$0.0535(35)$~fm$^2$~\cite{jap}}, and  $0.068(1)$~fm$^2$~\cite{arnold}. 
In Ref.~\cite{darms} the $(\gamma,n)$ cross section was deduced from the $(e,e')$ data as described above. 
Employing Eq.~(\ref{def}) and the resonant parameters of Ref.~\cite{darms} for a fit of the constant $C$ to the $(\gamma,n)$ 
cross section of  Ref.~\cite{darms}, the result is $B(E1)\!\!\uparrow=0.057$~fm$^2$ with the $\chi^2$ value per datum being 1.2. 
This $B(E1)\!\!\uparrow$ value obtained from  $(e,e')$ data is rather close to the quoted values of Refs.~\cite{sum,jap}. 
However, it differs noticeably from the HI$\gamma$S  $(\gamma,n)$--data result \cite{arnold}. 
One may note that the $(\gamma,n)$ cross section of Ref.~\cite{arnold} in the region near threshold  is larger than those 
of the other measurements but agrees with them at energies $E_\gamma > 1.9$~MeV (cf.\ Fig.~7 in Ref.~\cite{arnold}). 

Another possible definition of $B(E1)\!\!\uparrow$ is as follows. 
One chooses  $E_{max}$ in Eq.~(\re{BE1}) such that the integration covers the region where the 1/2$^+$ state dominates. 
In particular, one is forced to proceed in this way when contributions from other states are not separated out. 
Difference in the two definitions of the reduced strength may have led to the controversy in the literature.
For example, let us take as $E_{max}$ the last point, $E_\gamma=$2.2039 MeV, of the $(\gamma,n)$ cross section deduced in Ref.~\cite{darms} 
as a reasonable choice. 
The integration up to this $E_{max}$ value of the  $b(E1,E_\gamma)$ spectrum of Ref. \cite{darms},  obtained from the $(e,e')$ data, 
gives \mbox{$B(E1)\!\!\uparrow=0.034\, {\rm fm}^2$}. 
A $B(E1)\!\!\uparrow$ value of \mbox{$0.034(3)\, {\rm fm}^2$}  was also obtained in the $(e,e')$ experiment of Ref.~\cite{bates} 
although the $E_{max}$ value employed there was not specified. 
And the integration of the real photon $b(E1,E_\gamma)$ spectrum  of  Ref.~\cite{sum} up to this $E_{max}$ limit 
gives \mbox{$B(E1)\!\!\uparrow=0.033\, {\rm fm}^2$} close to the above value deduced from the $(e,e')$ experiments. 
The $B(E1)\!\!\uparrow$ values from the $(e,e')$ experiments reported in Refs.~\cite{darms,darms1} were 0.027(2)  fm$^2$. 
This would correspond to a somewhat lower value of $E_{max}$. 
However, the $B(E1)\!\!\uparrow$ strength extracted from the real photon  $b(E1,E_\gamma)$ spectrum of Ref.~\cite{arnold} would to be 0.044~fm$^2$ for an integration up to $E_{max}=$2.2039~MeV value. 

The big differences between the $B(E1)\!\!\uparrow$ values obtained in the two ways discussed above are 
caused by the fact that an integral of the type of Eq.~(\re{pi}) with a finite upper limit $E_{max}$ of integration 
converges very slowly as $E_{max}$ approaches to infinity. 
The contribution in Eq.~(\ref{BE1}) from excitation energies outside the region, where the resonance dominates the cross sections, is substantial.  
Such an extension of the integration far beyond the vicinity of a resonance has no direct physical meaning and is a matter of convention.

The conclusion thus is that the $B$(E1) strengths for the transition to the 1/2$^+$ state obtained from the $(e,e')$ \cite{darms, bates} 
and  $(\gamma,n)$ \cite{sum,jap} experiments agree well to each other in the above interpretation while they differ significantly 
from the value obtained in the latest  $(\gamma,n$)  experiment \cite{arnold}. 
While we cannot provide an explanation, we note that a similar kind of discrepancy exists for the prominent neutron spin--flip $M1$ transition in $^{48}$Ca whose strength plays a central role in understanding the phenomenon of quenching of the spin--isospin response \cite{heyde}.
It was first observed in electron scattering \cite{steffen} and recently remeasured \cite{tompkins} with the same experimental technique as used in Ref.~\cite{arnold}.
Similar to the $^9$Be case, the corresponding $1^+$ state lies close to the neutron threshold. 
The $(\gamma,n)$ cross sections and thus the $B(M1)\!\!\uparrow$ value deduced in Ref.~\cite{tompkins} is almost twice that obtained in Ref.~\cite{steffen}, while the cross sections at higher $\gamma$ energies agree with those from an $(e,e'n)$ experiment \cite{strauch}.

\section{Summary}

The present work addressed two questions concerning the structure of the first excited state of $^9$Be deduced from studies with electromagnetic probes.
First, all recent experimental data on the first excited state of $^9$Be testify to that this state is a resonance. 
Its energy and width are deduced, to be reproduced in forthcoming microscopic calculations of resonances as complex energy eigenstates. 
The energy value proved to be negative. 
Secondly, in difference to statements in the literature a fair agreement between the $B(E1)$ values obtained from the $(e,e')$ and one of the  $(\gamma,n)$ data sets can be found assuming different ways of extraction of the numbers in the literature.
The latest $(\gamma,n)$ measurement \cite{arnold} shows larger cross sections near threshold than all the other experiments which remain to be understood.  

\begin{acknowledgments}

VDE would like to thank ECT* for its kind hospitality during an extended stay, from which the present work originated.
We acknowledge a useful comment by H.~T.~Fortune.
This work was supported by the DFG under contract SFB 634 (PvNC and AR) and by RFBR grants \mbox{13--02--01139} and NS--215.2012.2 (VDE) .

\end{acknowledgments}


\begin{thebibliography}{99}
\bibitem{sasa}
T. Sasaqui, K.T. Kajino, G. Mathews, K. Otsuki, and T. Nakamura, Astrophys. J. {\bf 634}, 1173 (2005).
\bibitem{darms}O. Burda, P. von Neumann-Cosel, A. Richter, C. Forss\'en, and B.A. Brown, Phys. Rev. C {\bf 82}, 015808 (2010).
\bibitem{sum}K. Sumioshi, H. Utsunomiya, S. Goko, and T. Kajino, Nucl. Phys. A {\bf 709}, 467 (2002). 
\bibitem{arnold} C.W. Arnold, T.B. Clegg, C. Iliadis, H.J. Karwowski, G.C. Rich, J.R. Tompkins, and C.R. Howell,
Phys. Rev. C {\bf 85}, 044605 (2012).
\bibitem{jap}H. Utsunomiya, Y. Yonezawa, H. Akimune, T. Yamagata, M. Ohta, M. Fujishiro, H. Toyokawa, and H. Ohgaki,
Phys. Rev. C {\bf 63}, 018801 (2000).
\bibitem{arai}K. Arai, Y. Ogawa, Y. Suzuki, and K. Varga, Phys. Rev. C {\bf 54}, 132 (1996).
\bibitem{koike} Y. Koike, E. Cravo, and A.C. Fonseca, in {\it Proceedings of the International
Symposium on Clustering Aspects of Quantum Many--Body Systems, Kyoto, Japan, 2001}, edited by A. Ohnishi {\it et al.}
(World Scientific, Singapore, 2002), p. 65.
\bibitem{arai1}K. Arai, P. Descouvemont, D. Baye, and W.N. Catford, Phys. Rev. C {\bf 68} 014310 (2003).
\bibitem{Furutani}H. Furutani, H. Kanada, T. Kaneko, S. Nagata, H. Nishioka, S. Okabe, S. Saito, T. Sakuda, and
M. Seya, Progr. Theor. Phys. Suppl. {\bf 68}, 193 (1980).
\bibitem{gar} E. Garrido, D.V. Fedorov, and A.S. Jensen, Phys. Lett. B {\bf 684}, 132 (2010). 
\bibitem{dima}R. \'Alvarez-Rodr\'iguez, A.S. Jensen, E. Garrido, and D.V. Fedorov, Phys. Rev. C {\bf 82}, 034001 (2010).
\bibitem{bang}V.D. Efros and  J. Bang, Eur. Phys. J. A {\bf 4}, 33 (1999).
\bibitem{lud}G. L\"uders, Z. Naturforsch. {\bf 10a}, 581 (1955); see also, e.g., 
A.I. Baz', Y.B. Zeldovich, and A.M. Perelomov, {\it Scattering, Reactions and Decay in Nonrelativistic 
Quantum Mechanics} (Israel Program for Scientific Translations, Jerusalem, 1969), Chapt. III. 
\bibitem{clerc} H.-G. Clerc, K. Wetzel, and E. Spamer, Nucl. Phys. {\bf A120}, 441 (1968).
\bibitem{barker1} F.C. Barker, Can. J. Phys. {\bf 61}, 1371 (1983).
\bibitem{ajz} J.M. Eisenberg and W. Greiner, {\it Nuclear Theory Vol 2: Excitation Mechanisms of the Nucleus} (North Holland, 1970). 
\bibitem{bates}J.P. Glickman {\it et al.}, Phys. Rev. C {\bf 43}, 1740 (1991).
\bibitem{darms1}G. K\"uchler, A. Richter, and W. von Witsch, Z. Phys. A {\bf 326}, 447 (1987).

\bibitem{heyde}
K. Heyde, P. von Neumann-Cosel, and A. Richter, Rev. Mod. Phys. {\bf 82}, 2365 (2010).

\bibitem{steffen}
W. Steffen, H.-D. Gr\"af, W. Gross, D. Meuer, A. Richter, E. Spamer, O. Titze, and W. Kn\"upfer,  Phys. Lett. B {\bf 95}, 23 (1980).

\bibitem{tompkins}
J.R. Tompkins, C.W. Arnold, H.J. Karwowski, G.C. Rich, L.G. Sobotka, and C.R. Howell, Phys. Rev. C {\bf 84}, 044331 (2011).

\bibitem{strauch}
S. Strauch, P. von Neumann-Cosel, C. Rangacharyulu, A. Richter, G. Schrieder, K. Schweda, and J. Wambach, Phys. Rev. Lett. {\bf 85}, 2913 (2000).


\end{thebibliography}
\end{document}